\documentclass[floats,floatfix,showpacs,amssymb,prd,superscriptaddress,twocolumn,aps]{revtex4-1}
\usepackage{graphicx, epsfig, amssymb} 
\usepackage{amsmath, amsfonts}
\usepackage{bm} 

\usepackage[linktocpage]{hyperref}
\usepackage[caption=false]{subfig}
\usepackage[usenames]{color}

\def\be{\begin{equation}}
\def\ee{\end{equation}}
\def\beq{\begin{eqnarray}}
\def\eeq{\end{eqnarray}}

\newcommand{\bea}{\begin{eqnarray}}
\newcommand{\eea}{\end{eqnarray}}
\newcommand{\ben}{\begin{enumerate}}
\newcommand{\een}{\end{enumerate}}
\newcommand{\bi}{\begin{itemize}}
\newcommand{\ei}{\end{itemize}}

\newcommand{\nn}{\nonumber}

\begin{document}

\title{{\LARGE Proca Stars:} \\ {\large gravitating Bose-Einstein condensates of massive spin 1 particles}}
 \author{Richard Brito}
\affiliation{CENTRA, Departamento de F\'\i sica, Instituto Superior T\'ecnico -- IST, Universidade de Lisboa -- UL, Avenida Rovisco Pais 1, 1049, Lisboa, Portugal}
 \author{Vitor Cardoso}
 \affiliation{CENTRA, Departamento de F\'\i sica, Instituto Superior T\'ecnico -- IST, Universidade de Lisboa -- UL, Avenida Rovisco Pais 1, 1049, Lisboa, Portugal}
 \affiliation{Perimeter Institute for Theoretical Physics, Waterloo, Ontario N2J 2W9, Canada}
 \author{Carlos A. R. Herdeiro}
 \affiliation{Departamento de F\'isica da Universidade de Aveiro and CIDMA, 
 Campus de Santiago, 3810-183 Aveiro, Portugal}
 \author{Eugen Radu}
 \affiliation{Departamento de F\'isica da Universidade de Aveiro and CIDMA, 
 Campus de Santiago, 3810-183 Aveiro, Portugal}

\date{August 2015}

\begin{abstract}
We establish that massive  complex Abelian vector fields (mass $\mu$) can form gravitating solitons, when minimally coupled to Einstein's gravity. Such \textit{Proca stars} (PSs) have a stationary, everywhere regular and asymptotically flat geometry. The Proca field, however, possesses a harmonic time dependence (frequency $w$), realizing Wheeler's concept of \textit{geons} for an Abelian spin 1 field. We obtain PSs with both a spherically symmetric (static) and an axially symmetric (stationary) line element.  The latter form a countable number of families labelled by an integer $m\in \mathbb{Z}^+$. PSs, like (scalar) boson stars, carry a conserved Noether charge, and are akin to the latter in many ways. In particular, both types of stars exist for a limited range of frequencies and there is a maximal ADM mass, $M_{max}$, attained for an intermediate frequency. For spherically symmetric PSs (rotating PSs with $m=1,2,3$), $M_{max}\simeq 1.058 M_{Pl}^2/\mu$ ($M_{max}\simeq 1.568,\, 2.337, \, 3.247 \, M_{Pl}^2/\mu$),  
slightly larger values than those for (mini-)boson stars. We establish perturbative stability for a subset of solutions in the spherical case and anticipate a similar conclusion for fundamental modes in the rotating case. The discovery of PSs opens many avenues of research, reconsidering five decades of work on (scalar) boson stars, in particular as possible dark matter candidates.
\end{abstract}


\maketitle

\noindent{\bf{\em Introduction.}}
According to the latest cosmological data~\cite{Ade:2015xua}, $\sim26\%$ of the Universe's energy content is dark matter (DM). The fundamental nature of DM, however, is unknown. 
A widespread viewpoint is that DM consists of weakly interactive massive particles (WIMPs); popular candidates are the lightest supersymmetric particles, with masses $\gtrsim$ GeV, such as the neutralino~\cite{Jungman:1995df}. Despite its success in modeling structure formation, some shortcomings of the WIMPs DM model arise in small scales, such as the ``missing satellite" and the ``cuspy core" problems. Different proposals, some of which have been claimed to solve these problems (see~\cite{Suarez:2013iw,Li:2013nal} for reviews), introduce light or ultra-light bosonic particles/fields, with masses $\ll$ eV, which may form, gravitationally, macroscopic Bose-Einstein condensates. Even though such proposals have essentially a phenomenological character, ultra-light particles (axions) are a natural ingredient of the Peccei-Quinn mechanism to solve the strong CP problem~\cite{Peccei:1977hh} and may be motivated at a fundamental level by string theoretical constructions - the \textit{axiverse}~\cite{Arvanitaki:2009fg}. 

Gravitationally-bound bosonic structures are thus relevant in the context of DM searches.
The study of (scalar) Bose-Einstein condensates as DM candidates is often performed in a Newtonian limit. 
In the fully relativistic regime, such models yield gravitating solitons: (scalar) \textit{boson stars} (SBSs). These objects, initially proposed as a (scalar) realisation of Wheeler's \textit{geon} idea~\cite{Kaup:1968zz,Ruffini:1969qy}, have found a variety of applications,  from black hole mimickers in astrophysics, to particle models in TeV gravity scenarios~(see~\cite{Schunck:2003kk,Liebling:2012fv} for reviews). 

In light of recent proposals advocating massive spin 1 particles as a DM ingredient~\cite{Holdom:1985ag,ArkaniHamed:2008qn,Pospelov:2008jd,Goodsell:2009xc}, the study of the corresponding self-gravitating structures and their dynamics is of special importance~\footnote{In some DM literature, axion-light particles and hidden photons are now typically referred to as weakly interacting slim particles (WISPs)~\cite{Arias:2012az}.}.
In this letter we show that much like massive spin 0 particles, massive spin 1 particles can cluster as everywhere smooth, asymptotically flat lumps of energy under their own weight, producing gravitating solitons we dub \textit{Proca stars} (PSs). Moreover, we observe a close parallelism between the physical properties of PSs and SBSs.

\noindent{\bf{\em Einstein-(complex)-Proca theory.}}
We consider one complex Proca field, with mass $\mu$
(or equivalently, two real Proca fields with the same mass). 
It is described by the potential 1-form $\mathcal{A}$, and field strength $\mathcal{F}=d\mathcal{A}$. 
We denote the corresponding complex conjugates by an overbar, $\bar{\mathcal{A}}$ and $\bar{\mathcal{F}}$. The minimal Einstein-(complex)-Proca model is described by the action:
\begin{equation}
\mathcal{S}=\int d^4x \sqrt{-g}\left(\frac{R}{16\pi G}-\frac{1}{4}\mathcal{F}_{\alpha\beta}\bar{\mathcal{F}}^{\alpha\beta}-\frac{1}{2}\mu^2\mathcal{A}_\alpha\bar{\mathcal{A}}^\alpha\right) \ . \nonumber
\end{equation}
The Einstein and Proca field equations are, respectively,  
$G_{\alpha\beta}=8\pi G T_{\alpha \beta}$,
$\nabla_\alpha\mathcal{F}^{\alpha\beta}=\mu^2\mathcal{A}^\beta$, 
where the energy-momentum tensor reads:
\begin{eqnarray}
T_{\alpha\beta}=-\mathcal{F}_{\sigma(\alpha}\bar{\mathcal{F}}_{\beta)}^{\ \, \sigma}-\frac{1}{4}g_{\alpha\beta}\mathcal{F}_{\sigma\tau}\bar{\mathcal{F}}^{\sigma\tau}\ \ \ \ \  \nonumber  \\ +\mu^2\left[\mathcal{A}_{(\alpha}\bar{\mathcal{A}}_{\beta)}-\frac{1}{2}g_{\alpha\beta}\mathcal{A}_\sigma\bar{\mathcal{A}}^\sigma \right] \ . \nonumber
\end{eqnarray}
The Proca equations imply the Lorentz condition (which is not a gauge choice, but a dynamical requirement):
\begin{equation}
\nabla_\alpha\mathcal{A}^\alpha= 0 \ . \nonumber
\end{equation}

The global $U(1)$ invariance of the action, under the transformation $\mathcal{A}_\mu\rightarrow e^{i\alpha}\mathcal{A}_\mu$, with $\alpha$ constant, implies the existence of a conserved 4-current, $
j^\alpha=\frac{i}{2}\left[\bar{\mathcal{F}}^{\alpha \beta}\mathcal{A}_\beta-\mathcal{F}^{\alpha\beta}\bar{\mathcal{A}}_\beta\right]$.
From the Proca equation, $\nabla_\alpha j^\alpha=0$. Thus a conserved Noether charge exists, $Q$, obtained integrating the temporal component of the 4-current on a space-like slice $\Sigma$:
$Q=\int_\Sigma d^3x \sqrt{-g} j^t$.

\noindent{\bf{\em Spherically symmetric PSs.}}
%
We first consider spherically symmetric solutions, taking the following line element form
\begin{equation}
ds^2=-\sigma^2(r)N(r)dt^2+\frac{dr^2}{N(r)}+r^2d\Omega_2 \ , 
\label{ansatz1}
\end{equation}
where $N(r)\equiv 1-{2m(r)}/{r}$ and the Proca potential ansatz 
\begin{equation}
\mathcal{A}=e^{-iwt}\left[f(r)dt+ig(r)dr\right] \ .
\label{ansatz2}
\end{equation}
$\sigma(r),m(r),f(r),g(r)$ are all real functions of the radial coordinate only and $w$ is a real frequency parameter. As for SBSs the harmonic time dependence of $\mathcal{A}$ is crucial and the complex nature of $\mathcal{A}$ makes it compatible with the time-independent line element. But unlike SBSs there are now two independent radial functions for the `matter' field, making the analysis more involved (even more so in the rotating case).
 The Proca equations yield 
 \begin{equation}
 \frac{d}{dr}\left\{\frac{r^2[f'-wg]}{\sigma}\right\}=\frac{\mu^2r^2f}{\sigma N} \ , \qquad  wg-f'=\frac{\mu^2\sigma^2 N g }{w} \ , \nonumber
 \end{equation}
 where $'$ denotes radial derivative. These two equations imply the Lorentz condition constraint (which determines $f(r)$ in terms of the remaining functions). The essential Einstein equations read
\begin{eqnarray}
&&
m'=4\pi G r^2
\left[
\frac{(f'-wg)^2}{2\sigma^2}
+\frac{1}{2}\mu^2 \left(g^2N+\frac{f^2}{N\sigma^2}\right)
\right],
\nonumber \\
&&\frac{\sigma'}{\sigma}=4\pi G r  \mu^2
\left(g^2+\frac{f^2}{N^2\sigma^2} \right) \ . \nonumber
\end{eqnarray}

 For the ansatz (\ref{ansatz1})-(\ref{ansatz2}), the
Noether charge reads 
%
$Q=\frac{4\pi \mu^2}{w}\int^{\infty}_0dr\, r^2g(r)^2\sigma(r)N(r)$,
and the energy density measured by a static observer is
\begin{equation}
-T^t_t=
 \frac{(f'-wg)^2}{2\sigma^2}
+\frac{1}{2}\mu^2 \left(g^2N+\frac{f^2}{N\sigma^2} \right)\ . \nonumber
\end{equation}
Finally,  PSs satisfy the virial relation:
\begin{eqnarray}
 \int_0^\infty dr\, r^2\sigma 
 \left[
 \mu^2
 \left(
 g^2-\frac{f^2(4N-1)}{\sigma^2 N^2}
 \right)-\frac{(wg-f')^2}{\sigma^2}
 \right ]=0, \nonumber
\end{eqnarray}
used to test the numerical accuracy of the results. This relation can also be used to rule out non-gravitating solutions, $i.e$ `Proca-balls' without backreaction~\footnote{It was recently shown that self-interactions can, however, allow such non-backreacting `Proca-balls'~\cite{Loginov:2015rya}, in analogy with the Minkowski space $Q$-balls found in the scalar case~\cite{Coleman:1985ki}.}.

\noindent{\bf{\em Asymptotic behaviours.}}
An analysis of the field equations both near the origin and at spatial infinity reveals smooth behaviours.
Close to $r=0$ we find
\begin{eqnarray}
&&
f(r)=f_0+\frac{f_0}{6}\left(\mu^2-\frac{w^2}{\sigma_0^2}\right)r^2+\mathcal{O}(r^4)\ , \nonumber \\
&& g(r)=-\frac{f_0w}{3\sigma_0^2}r+\mathcal{O}(r^3) \ , \nonumber
\\
\nonumber
&&
m(r)=\frac{4\pi G f_0^2\mu^2}{6 \sigma_0^2}r^3+\mathcal{O}(r^5)\ , \\
&& 
\sigma(r)=\sigma_0+\frac{4\pi G f_0^2 \mu^2}{2\sigma_0}r^2+\mathcal{O}(r^4) \ , \nonumber
\end{eqnarray}
where $f_0$, $\sigma_0$ are constants, the values of $f(r)$ and $\sigma(r)$ at the origin, respectively.
As $r\to \infty$, we find
\begin{eqnarray}
&&
f(r)= c_0\frac{e^{-r\sqrt{\mu^2-w^2}}}{r}+\dots\ , \nonumber \\
&& 
g(r)=c_0\frac{w}{\sqrt{\mu^2-w^2}}\frac{e^{-r\sqrt{\mu^2-w^2}}}{r}+\dots \ , \nonumber 
\\
\nonumber
&&
m(r)=M+\dots, \\
&&
\log \sigma(r)= -4\pi G \frac{c_0^2 \mu^2}{2(\mu^2-w^2)^{3/2}}\frac{e^{-2r\sqrt{\mu^2-w^2}}}{r}+\dots\ , \nonumber
\end{eqnarray}
where $M$ is the ADM mass and $c_0$ is a constant; observe that $w<\mu$, which is a bound state condition.
%

\noindent{\bf{\em Numerical results.}}
The solutions that smoothly interpolate between the two above asymptotic behaviours are found numerically.
In numerics -- and in the following -- we set $\mu=1$, $4\pi G=1$,
by using a scaled radial coordinate $r\to r \mu$ (together with $m\to m \mu$, $w\to w/ \mu$)
and scaled potentials
$f\to f \sqrt{4\pi G}$, $g\to g \sqrt{4\pi G}$. 
The equations are solved by using a standard Runge-Kutta ODE solver
and implementing a shooting method in terms of the parameter $f(0)$.

\begin{figure}[h!]
\centering
\includegraphics[height=2.15in]{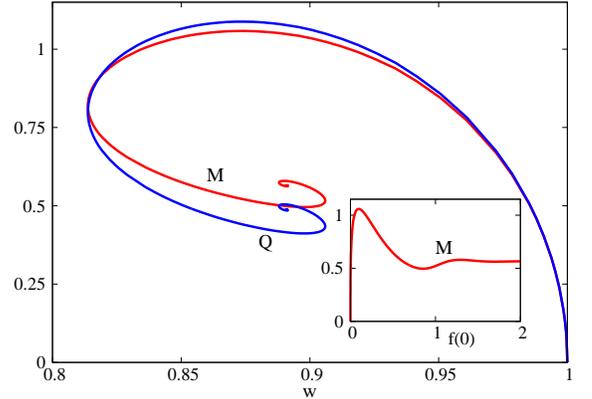}  
\caption{ ADM mass $M$ and the Noether charge $Q$ for spherical PSs $vs.$ the Proca field frequency $w$. The inset shows $M$ $vs.$ the shooting parameter $f(0)$, which is analogous to the curve for $M$ $vs.$ the central value of the scalar field for SBSs.
}
\label{spirals}
\end{figure}

In Fig.~\ref{spirals} we plot the ADM mass, $M$, and the Noether charge, $Q$, as a function of the scalar field frequency, $w$. 
As $w\to 1$, the mass and Noether charge of the solutions vanish, but $Q/M\rightarrow 1$. Approaching this limit, PSs -- like SBSs -- become spatially diluted with an effective size much larger than their Schwarzschild radii~\footnote{For a discussion on a possible definition of an effective radius for SBSs see, $e.g.$~\cite{Herdeiro:2015gia}.}, and trivialize at $w=1$. Reducing $w$ from this maximal value, PSs become more compact; both $M$ and $Q$ follow a spiral, towards a central critical configuration located around $w \simeq 0.891$. The maximum of both $M$ and $Q$ occurs for $w_{max}\simeq 0.875$, with $M_{max}\simeq 1.058<Q_{max}\simeq 1.088$. We observe that $M< Q$ from $w=1$ down to almost the minimal allowed frequency. In the lower part of the spiral $M> Q$, and the binding energy $1-M/ Q$ is negative, corresponding to a region of ``energy excess", where the solutions must be unstable against perturbations. All these features follow closely those of spherical SBSs.

\begin{figure}[h!]
\centering
\includegraphics[height=2.15in]{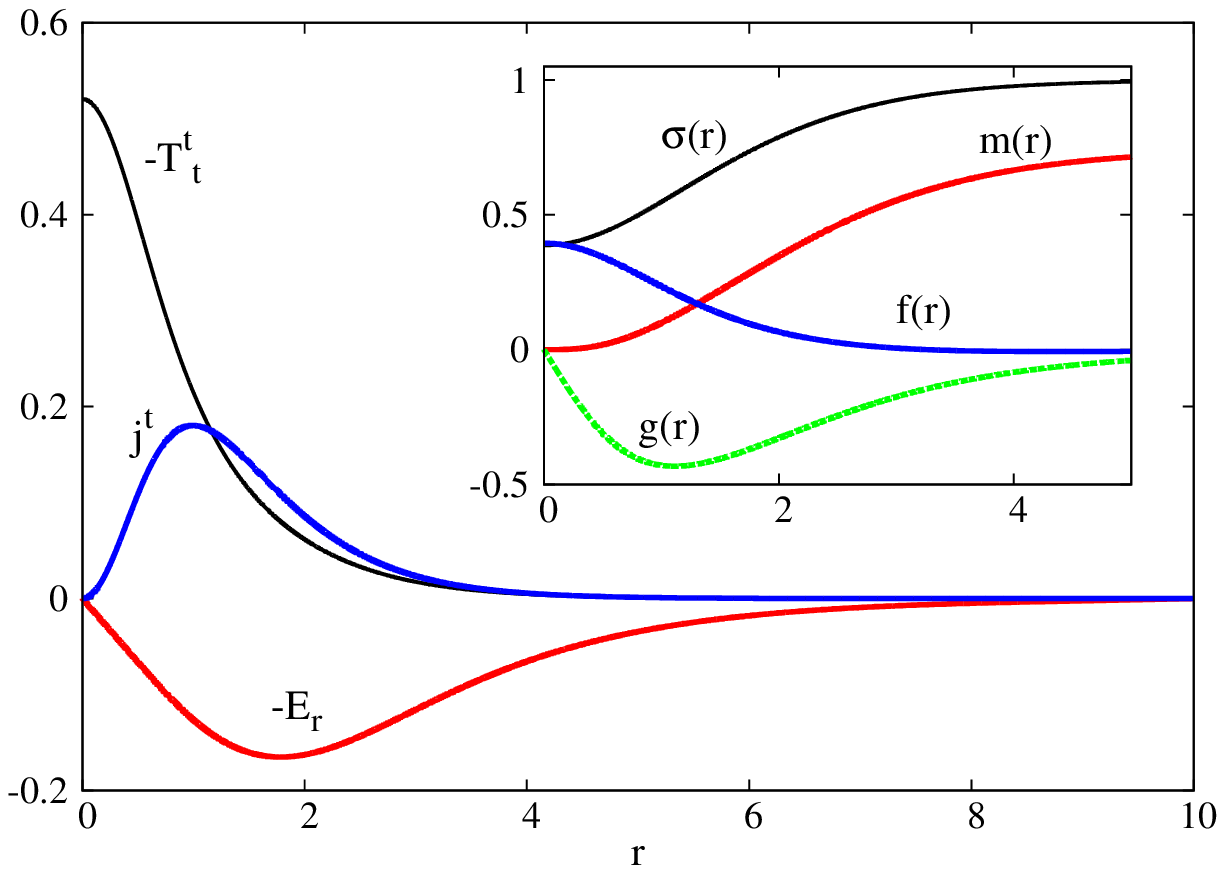}  
\includegraphics[height=2.15in]{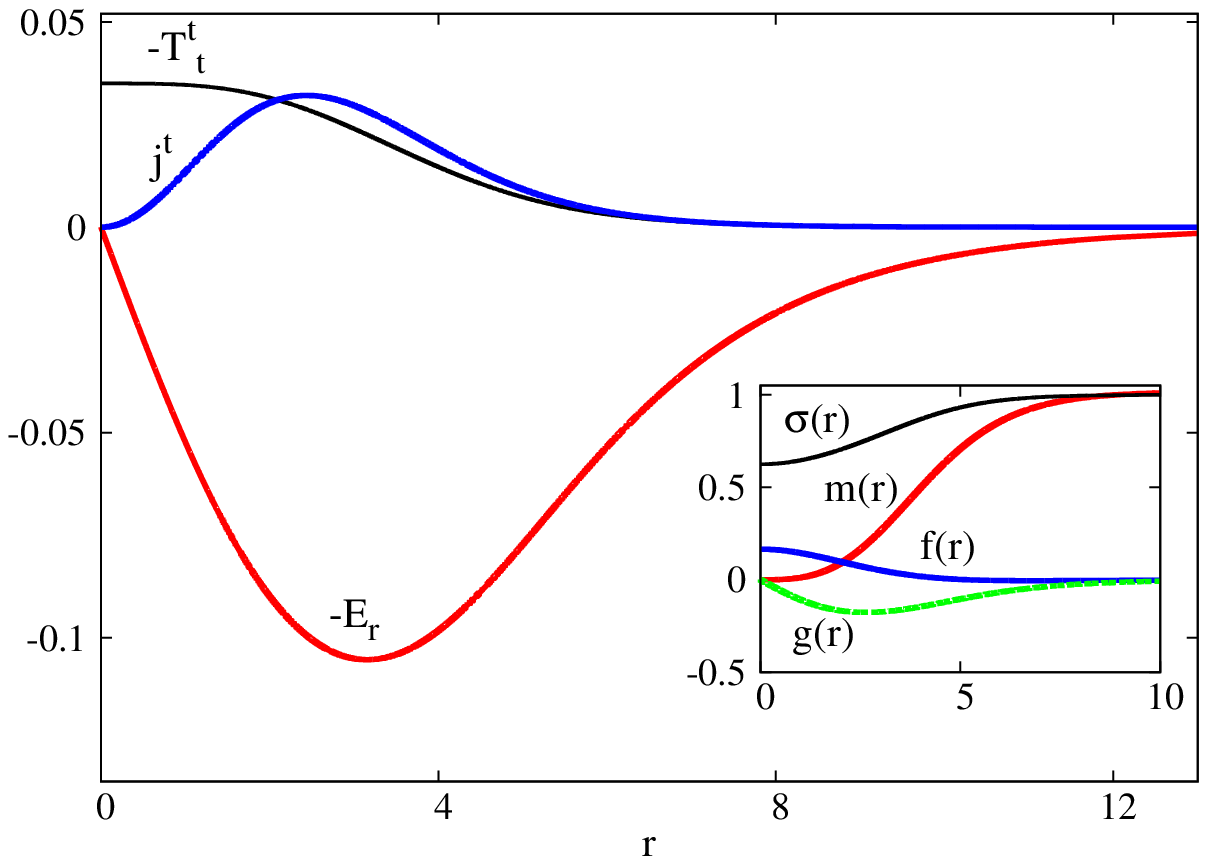}  
\caption{Energy and Noether charge densities, electric field ($E_r=|F_{rt}|$) and profile functions (inset) of two PSs solutions with $M=0.745, w=0.817, f(0)=0.394$ (top panel) and $M= 1.016, w= 0.839, f(0)=0.165$ (lower panel). 
}
\label{profile}
\end{figure}

We remark that for the family of solutions exhibited in Fig.~\ref{spirals}, $f(r)$ has one node~\footnote{The existence of a node is not obvious from the plots;  but one can show analytically that, for any spherical PS, $f(r)$ changes sign at least once. This differs from SBSs, for which fundamental modes are nodeless in the (single) profile function.} and $g(r)$
is nodeless. These are the fundamental solutions. There are also excited solutions, with more nodes for these functions, but we did not attempt to study them in any systematic way. Studies of SBSs suggest that excited solutions are unstable~\cite{Balakrishna:1997ej}. 
In Fig.~\ref{profile} we show the behaviour of physical and profile functions for two PS solutions. In particular observe the energy concentration near $r=0$.

\noindent{\bf{\em Stability.}}
A positive binding energy -- as that observed along the upper branch of Fig.~\ref{spirals} -- does not guarantee linear stability. For SBSs, a perturbative stability analysis shows that the solutions are only stable from the maximal frequency $w=1$ down until the frequency at which the maximal mass is attained, $w_{max}$. Beyond this point an unstable mode develops~\cite{Gleiser:1988ih,Lee:1988av}. An analogous conclusion holds for PSs, as we now show by considering their linear radial perturbations.

\begin{figure}[htb]
\begin{center}
\begin{tabular}{c}
\includegraphics[height=2.15in]{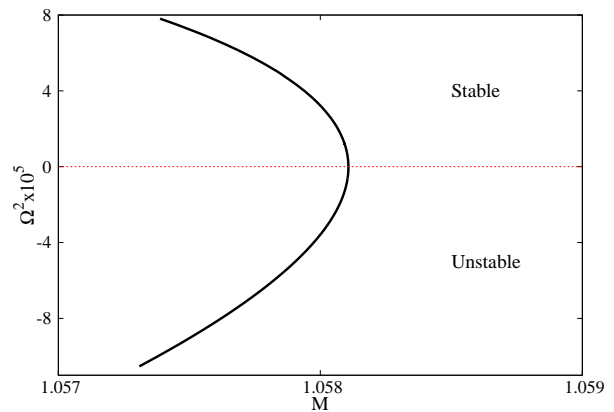}  
\end{tabular}
\caption{Fourier frequency $\Omega$ of the perturbations as a functions of the PS's total mass. The critical mass at which PSs become unstable corresponds to the star's maximum mass.\label{stability}}
\end{center}
\end{figure}

We assume that all perturbations have a harmonic time dependence of the form $e^{-i \Omega t}$, with $\Omega$ being the characteristic vibrational frequencies of the PS. Gauge freedom allows to write the perturbed metric as:
\beq\label{metric}
ds^2 =&& -\sigma^2(r)N(r)\left[1- \epsilon h_0(r)e^{-i \Omega t} \right]dt^2\nn\\
&&+\frac{dr^2\left[1+ \epsilon h_1(r) e^{-i \Omega t} \right]}{N(r)}+r^2d\Omega^2\, ;
\eeq
the vector field is perturbed as
\beq\label{vector}
\mathcal{A}= && e^{-i w t}\left[\left(f(r)+ e^{-i \Omega t}\frac{ \epsilon f_1(r)+i\epsilon f_2(r)}{r}\right) dt\right.\nn\\
&&+\left.\left(i g(r)+e^{-i \Omega t}\frac{\epsilon g_1(r)+i \epsilon g_2(r)}{r}\right) dr\right]\,,
\eeq
where $h_0$, $h_1$, $f_1$, $f_2$, $g_1$ and $g_2$ are radial perturbations around the background solution, and $\epsilon$
is a small, bookkeeping parameter. The perturbed Einstein-Proca system can be reduced to a pair of second order ODE's. Imposing regularity of the perturbations at the origin and at infinity, the resulting system is a two dimensional eigenvalue problem for $\Omega$ and one other constant which we have chosen to be the value of $h_0$ at the origin. A numerical solution is then obtained by a two dimensional shooting, with the result shown in Fig.~\ref{stability}, where we plot $\Omega^2$ versus the PS's ADM mass around the maximal mass. 
For $\Omega^2>0$ the mode is purely real and corresponds to stable normal modes. This occurs for $w>w_{max}$. For $\Omega^2<0$ the mode is a purely positive imaginary number, indicating that these configurations are unstable ($cf.$ Eqs.~\eqref{metric} and \eqref{vector}). Thus, in complete agreement with the SBSs case~\cite{Gleiser:1988ih,Lee:1988av,Seidel:1990jh,Hawley:2002zn}, we can see that the star's maximum mass corresponds to a transition point between stable and unstable configurations.  
For unstable configurations with mass $0.1\%$ below the threshold, the instability timescale $\tau$ is already smaller than $\tau \lesssim 100 M$. Thus, we expect the unstable branch of PS to share similar properties to those observed in the scalar case:
configurations which reach this branch will either quickly collapse to black holes or migrate back to the stable branch via mass ejection (the ``gravitational cooling'' mechanism)~\cite{Seidel:1990jh,Seidel:1993zk,Alcubierre:2003sx,Brito:2015yga}.

\noindent{\bf{\em Rotating PSs.}}
We now turn to rotating PSs. These solutions are found with an axi-symmetric metric ansatz
\beq
  ds^2=-&& e^{2F_0(r,\theta)}dt^2+e^{2F_1(r,\theta)}(dr^2+r^2 d\theta^2) \nn\\
 && +e^{2F_2(r,\theta)}r^2 \sin^2\theta \left(d\varphi-\frac{W(r,\theta)}{r}dt\right)^2 \ . \nonumber
\eeq

The Proca field ansatz is given in terms of another four functions $(H_i,V)$ which depend also on $r,\theta$
\begin{eqnarray}
A=\left(
\frac{H_1}{r}dr+H_2d\theta+i H_3 \sin \theta d\varphi + iVdt   
\right)
e^{i(m\varphi-w t)} \, , \ \ \ \ \ \nonumber
\end{eqnarray}
 with $m\in \mathbb{Z}^+$. The Einstein-Proca equations are solved with the following boundary conditions, which we have found to be compatible with an approximate construction of the solutions
on the boundary of the domain of integration: $\partial_r F_i\big|_{r=0}=W\big|_{r=0}=H_i|_{r=0}=V|_{r=0}=0$; also $ F_i\big|_{r=\infty}=W\big|_{r=\infty}=H_i|_{r=\infty}=V|_{r=\infty}=0$; and $ \partial_\theta F_i\big|_{\theta=0,\pi}=\partial_\theta W\big|_{\theta=0,\pi}=H_i|_{\theta=0,\pi}
=V|_{\theta=0,\pi}=0$
(note that for $m=1$ the conditions for  $H_2,  H_3$ are different, with 
$\partial_\theta H_2\big|_{\theta=0,\pi}=\partial_\theta H_3\big|_{\theta=0,\pi}=0$).
All rotating PSs we have constructed so far are symmetric $w.r.t.$ a reflection along the equatorial plane.
Odd-parity composite configurations, however, are also likely to exist.

As usual, the ADM mass $M$ and angular momentum $J$ are read off from the asymptotic expansion,
$g_{tt} =-1+{2GM}/{r}+\dots,~~g_{\varphi t}=-{2GJ}\sin^2\theta/r+\dots$. 
In analogy with the SBSs case~\cite{Yoshida:1997qf}, one can show that the angular momentum
is a multiple of the Noether charge, $J = m Q$, 
but in contrast to the SBSs case, the angular momentum density $T_\varphi^t$ and 
the Noether charge density $j^t$ are not proportional any longer.

\begin{figure}[h!]
\centering
\includegraphics[height=2.09in]{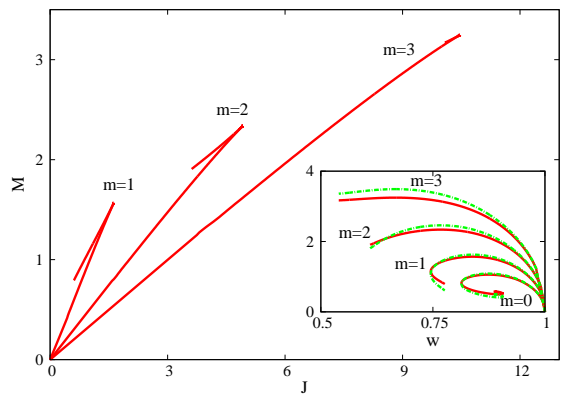}  
\caption{$M$ $vs.$ $J$ diagram for  rotating PSs with $m=1,2,3$. The inset shows  $M$ (solid red lines) and $Q$ (dashed green lines) $vs.$ $w$, where  $m=0$ corresponds to the spherical PSs in Fig.~\ref{spirals}.}
\label{rotating}
\end{figure}

The Einstein-Proca equations for this rotating case are quite involved and shall not be presented here. They are solved numerically, subject to the above boundary conditions, by 
using the elliptic PDE solver FIDISOL~\cite{schoen} based on the Newton-Raphson procedure, employing a compactified radial coordinate $x=r/(1+r)$~\footnote{We use a scaling similar to that described in the spherical case. We estimate the typical error of the solutions to be smaller than 1 part in $10^4$. In the spherical case much higher accuracies are obtained; $e.g.$, the virial identity is typically obeyed to 1 part in $10^{12}$.}. In Fig.~\ref{rotating} we exhibit the mass $vs.$ angular momentum, together with the  mass and the Noether charge $vs.$ the Proca field frequency, for rotating PSs with $m=1,2,3$ (inset). Both panels exhibit the familiar shape from the study of SBSs. In particular $M$ and $J$ are always positively correlated~\footnote{With increasing $m$ it becomes increasingly difficult, numerically, to move further along the spiral.}. The physical quantities for the largest mass PSs are exhibited in the following table, for $m=0$ (spherical) and $m=1,2,3$ (rotating), where they are compared with the results in the literature for the corresponding largest mass SBSs (last column); one observes that $M_{max}$ is always smaller for the latter.

{\small \begin{center}
\begin{tabular}{c|c|c|c|c||c}
\hline
 &   $ \ \ w_{max}$ & \ \ $M_{max}$ & $Q_{max}$ & $J_{max}$ & $M_{max}$ SBSs \\
\hline
$m=0$ \ \   & \ \     0.875 \ \ & \ \ 1.058  \ \ & \ \  1.088 \ \ & \ \ 0 \ \ & \ \ 0.633~\cite{Kaup:1968zz} \\
 $m=1$  \ \  & \  \  0.839 \ \ & \ \ 1.568 \ \ & \ \ 1.626 \ \  & \ \ 1.626 \ \ & \ \ 1.315~\cite{Yoshida:1997qf,Grandclement:2014msa}  \\
 $m=2$ \ \  &   \ \  0.767 \ \ & \ \ 2.337 \ \ & \ \ 2.461 \ \ & \ \ 4.921 \ \ & \ \ 2.216~\cite{Grandclement:2014msa}  \\
 $m=3$ \ \  &   \ \  0.667 \ \ & \ \ 3.247 \ \ & \ \ 3.489 \ \ & \ \ 10.467 \ \ & \ \ {\rm not \ reported} \\
 \hline
\end{tabular}\\
\bigskip
\end{center}}

%

Other physical properties of the rotating PSs mimic those of SBS: the rotating PSs solutions spatially delocalize as $w\to 1$ (similarly to the static case) and trivialize in that limit;  
the energy density has an essentially toroidal distribution;
also, for $m=1$, neither the energy density nor the Noether charge density vanish on the symmetry axis~\footnote{For PSs, however, extra structure appears that will be discussed in detail elsewhere.}.

We have not attempted to study in detail the stability of rotating PSs. For rotating SBSs, catastrophe theory arguments 
\cite{Kleihaus:2011sx} support a similar conclusion to that obtained in the static case: the solutions are stable from the maximal frequency down to the point where the maximal mass is attained. 
We expect a similar conclusion to hold for rotating PSs. In the rotating case there can also occur ergoregion instabilities~\cite{Brito:2015oca}; for SBSs ergoregions only appear in the unstable branch of solutions and an analogous situation is true for PSs~\footnote{For $m=1,2,3$ an ergo-region -- actually an ergo-torus -- arises on the first branch for $w=0.745, 0.671, 0.648$, respectively, thus for $w$ smaller than the corresponding $w_{max}$.}.

\noindent{\bf{\em Discussion and Outlook.}}
The discovery of PSs suggests various novel directions of research, besides the detailed analysis of the solutions presented herein; let us mention two. $1)$ In the scalar case, a single (rather than complex) real field allows the existence of very long lived quasi-solitons -- \textit{oscillatons}; the same is true for the Proca case ~\cite{Brito:2015yga}. $2)$ Can one add a black hole inside a PS, like for other gravitating solitons? While for spherically symmetric configurations one can show the absence of black hole solutions, this is possible inside a spinning PSs, again in complete analogy with the SBSs case~\cite{Herdeiro:2014goa,Herdeiro:2014ima,Herdeiro:2015gia}.

A final remark relating PSs with DM phenomenology. Since ordinary matter, which constitutes only $\sim 5\%$ of the Universe's energy content, is comprised of over ten elementary particles, it is reasonable to admit that DM is composed of different kinds of fundamental entities, but which, when gravitationally clustered into macroscopic lumps, display some universality.  If  Bose-Einstein condensates of ultra-light scalar fields are viable DM models, the existence of PSs akin to SBSs suggests the former may be another DM component. In this context it would be interesting to study the Newtonian limit of PSs.

\newpage

\noindent{\bf{\em Acknowledgements.}} 
R.B. acknowledges financial support from the FCT-IDPASC program through the grant SFRH/BD/52047/2012, and from the Funda\c c\~ao Calouste Gulbenkian through the Programa Gulbenkian de Est\' imulo \`a Investiga\c c\~ao Cient\'ifica.
V.C. acknowledges financial support provided under the European
Union's FP7 ERC Starting Grant ``The dynamics of black holes: testing
the limits of Einstein's theory'' grant agreement no. DyBHo--256667,
and H2020 ERC Consolidator Grant ``Matter and strong-field gravity: New frontiers in Einstein's theory'' grant agreement no. MaGRaTh--646597. 
C.H. and E.R. thank funding from the FCT-IF programme and the grants PTDC/FIS/116625/2010, NRHEP--295189-FP7-PEOPLE-2011-IRSES.
This research was supported in part by the Perimeter Institute for
Theoretical Physics, and by the Government of Canada through Industry Canada and by the Province
of Ontario through the Ministry of Economic Development $\&$
Innovation.
%


\bibliographystyle{h-physrev4}
\bibliography{letter}
\end{document}